# Experimental Measurements of Benzene Ice Critical Saturation Ratios on Titan Tholins and Impact on the Microphysical Modeling of Titan's South Polar Benzene Cloud

David Dubois[1,2,3], Erika Barth[4], Laura T. Iraci[1], Ella Sciamma-O'Brien[1], Farid Salama[1], Sandrine Vinatier[5]

## Abstract

The Cassini Composite Infrared Spectrometer (CIRS) revealed the presence of a benzene ($C_6H_6$) ice cloud in Titan's autumn south polar stratosphere, following the northern spring equinox in August 2009. This event increased the mixing ratio of benzene and raised the cloud top altitude. We present here new experimental measurements of the critical saturation ($S_{crit}$) of pure $C_6H_6$ ice at low temperature (from 13.8 at 138 K to 3.3 at 157 K), representative of Titan's atmospheric temperatures, when deposited onto Titan aerosol analogs (*tholin*s) produced in the NASA Ames COsmic SImulation Chamber (COSmIC). Comparisons of $S_{crit}$ values of benzene vapor deposition obtained on blank substrates versus on Titan *tholins* reveals the highly favorable role aerosols play as condensation nuclei. We have included these new $S_{crit}$ measurements to calculate the nucleation contact parameter from $m = 2.6\text{e-}4\ T + 0.9494$ in the Community Aerosol and Radiation Model for Atmospheres (CARMA) to investigate the variation in size and number density of $C_6H_6$ cloud particles as a function of altitude in Titan's southern polar atmosphere at 87° S. We discuss how these new temperature-dependent measurements impact the microphysical modeling and result in simulated benzene clouds forming about 20 km lower than seen in previous observations. A possible explanation for this discrepancy is that the cloud system was likely influenced by the co-condensation of more than one volatile gas species.

[1] NASA Ames Research Center, Moffett Field, CA, USA
[2] Bay Area Environmental Research Institute, Moffett Field, CA, USA
[3] Present address: Centre de Recherche sur les Ions, les Matériaux et la Photonique Université Caen Normandie, ENSICAEN, CNRS, CEA, Normandie Univ, CIMAP UMR6252, F-14000 Caen, France
[4] Southwest Research Institute, Boulder, CO, USA
[5] LIRA, Observatoire de Paris, Université PSL, CNRS, Sorbonne Université, Université Paris Cité, Meudon, France

david.dubois@unicaen.fr

# 1. Introduction

Titan's circulation dynamics underwent a global change following the northern hemisphere spring equinox in 2009. The descending branch of the global circulation cell, observed in the south pole in 2010 resulted in a heating of the stratopause (at ~400 km) and an enrichment of haze at high southern latitude (>80º S, Teanby et al. 2012; Vinatier et al. 2015). This atmospheric circulation reversal was in agreement with general circulation model (GCM) predictions (*e.g.*, Lebonnois et al. 2012; Lora et al. 2015; Vatant d'Ollone et al. 2020) and has since been reanalyzed with updated GCMs (de Batz de Trenquelléon et al. 2025). The descending branch carried chemically-enriched gases originating from the upper atmosphere where the first photochemically-produced molecules are formed, as revealed by Cassini measurements which showed increased mixing ratios for many of these molecules (*e.g.*, $C_2H_2$, $C_2H_4$, $C_6H_6$, and HCN) above 400 km (Teanby et al. 2012; Vinatier et al. 2015; Coustenis et al. 2016; Mathé et al. 2020). In 2012, thermal cooling between 250 and 500 km occurred at the south pole (Vinatier et al. 2015; Teanby et al. 2017). This cooling was predominantly a direct consequence of radiative cooling caused by the organics-enriched descending branch. The thermal conditions thus became appropriate for several of these molecules to condense at high altitude (>250 km) in 2012. Analyses of Cassini observations by de Kok et al. (2014) and Vinatier et al. (2018) led to firm detections of HCN and $C_6H_6$ ice spectral signatures in the south polar cloud. More recently, the dynamics and evolution of the polar HCN cloud was investigated by Hanson et al. (2023, 2025) using microphysics simulations and analysis of Cassini/ISS images. This reanalysis of Cassini data revealed unexpected altitude and latitude variations of this cloud (also called high-altitude south polar cloud, or HASP) with a drop in altitude from 300 km down to 200 km over the course of the southern fall. For a more complete and recent description of these dynamics, the reader is referred to Figure 1 from de Batz de Trenquelléon (2026) and de Batz de Trenquelléon et al. (2026b).

Microphysics models have been used to characterize the ice nucleation and condensation mechanisms within the compositional and thermal conditions of Titan's stratosphere (*e.g.,* Barth 2017, Hanson et al. 2023) and were also coupled within GCM (de Batz de Trenquelléon et al. 2025b). The most recent study was able to simulate the altitude of the top of the southern polar cloud initially extending from 336 km with and showing a gradually decreasing altitude during the autumn, in agreement with Cassini/ISS observations (Hanson et al. 2025). When considering 100% humidity condensation curves to model Cassini CIRS-derived temperature profiles, models have been able to predict approximate maximum condensation altitudes of multiple organic species (*e.g.*, $C_6H_6$, HCN, $C_2H_2$, etc) based on key laboratory and theoretical microphysical constraints (Anderson & Samuelson, 2011; Barth 2017; Hanson et al. **2023**). Microphysics model parameters include the temperature-dependent equilibrium vapor pressure and critical saturation ratios $S_{crit}$ (a temperature-dependent value determined by the ratio between the nucleation partial pressure, *i.e.*, when supersaturation is reached, and the equilibrium vapor pressure) of species expected to condense. In addition, knowing $S_{crit}$ enables the calculation of the contact parameter *m* (*i.e.*, the cosine of the angle typically situated between a droplet and its underlying substrate) which itself determines the nucleation rate (Barth 2017). However, many of these parameters are often only valid in temperature regimes outside of Titan's cold stratosphere and require extrapolations from published measurements made at warmer temperatures. Such extrapolations can induce errors on the retrieved gas abundances and ice cloud properties (*e.g.*, Dubois et al. 2021). The paucity of available ice nucleation data measured in the laboratory at cold temperatures and the need to rely on extrapolation have therefore limited more realistic modeling of organic ice cloud observations on Titan. Addressing this need for laboratory measurements at relevant temperatures was the primary motivation for our previous study of $C_6H_6$ equilibrium vapor pressure (Dubois et al. 2021) and the study presented here, given the importance of many ice signatures (*e.g.*, HCN, $HC_3N$, $C_2N_2$) detected in Titan's stratospheric clouds at temperatures below 180 K (Anderson et al. 2018).

To further our study of $C_6H_6$ ice nucleation, we carried out experimental $S_{crit}$ measurements for $C_6H_6$ ice deposited between 138–157 K under vacuum (temperature conditions approximately corresponding to an altitude of 300 km), either on a bare silicon substrate or on Titan aerosol analogues called *tholins*. While our measurements on bare silicon provide a benchmark for benzene $S_{crit}$, the ice deposition experiments onto *tholins* aim to provide more representative conditions under which $C_6H_6$ ice nucleates and grows on the haze particles present in Titan's stratosphere. It is important to note that the physical and chemical properties of *tholins* are dependent upon the experimental techniques used to generate them (Cable et al. 2012). The size of the grains can range from tens of nm up to hundreds of nm (*e.g.*, Hadamcik et al. 2009; Sciamma-O'Brien et al. 2017), and their mechanical and optical properties can be directly linked to their formation parameters such as gas pressure, temperature, and energy source (*e.g.*, Li et al. 2022; Drant et al. 2026; Husić et al. 2026). Tholins can also be formed as amorphous grains that can aggregate (*e.g.*, Szopa et al. 2006; Hadamcik et al. 2009; Sciamma-O'Brien et al. 2017) or as thin films (*e.g.*, Sekine et al. 2008; Quirico et al. 2008), which have different bulk properties (*e.g.*, Cable et al. 2012; Hörst et al. 2013; Husić et al. 2026). The *tholins* we used in this study were amorphous grains formed in volume by plasma chemistry and jet-deposited on silicon substrates forming a film. After measuring the new laboratory $S_{crit}$ measurements, we then included them in the Community Aerosol and Radiation Model for Atmospheres (CARMA, Barth 2017) and derived the benzene mixing ratios and the cloud number density at 87° S in order to conduct a new analysis of the south polar cloud microphysics. Below we describe the experimental protocol, present the results, and discuss the microphysical impact of *tholins* on benzene condensation and its implications on the formation and evolution of Titan's high altitude south polar cloud system.

## 2. Methods

### 2.1 Experimental

#### 2.1.1. Titan Aerosol Analog Production

***Experimental Setup***

Titan *tholins* were produced in the Ames COSmIC facility, used to simulate Titan's atmospheric chemistry. In COSmIC, the chemistry is induced by a plasma discharge generated in the stream of a jet-cooled (150 K) planar expansion of $N_2$:$CH_4$ (95:5) gas. The COSmIC facility uses a pulsed discharge nozzle (PDN, Figure 1) to cool gas mixtures down to ~150 K and reduce the pressure to 30 mbar by expanding the gas through a 127 μm slit (Broks et al. 2005; Biennier et al. 2006; Sciamma-O'Brien et al. 2014; 2017). A plasma discharge is then generated, in the stream of the jet-cooled expansion, by applying a negative high voltage (-700 V to -1000 V) onto electrodes placed along the slit. The plasma is the energy source that induces chemistry in the cooled gas and results in the formation of larger molecules and solid particles, *i.e.*, *tholins*.

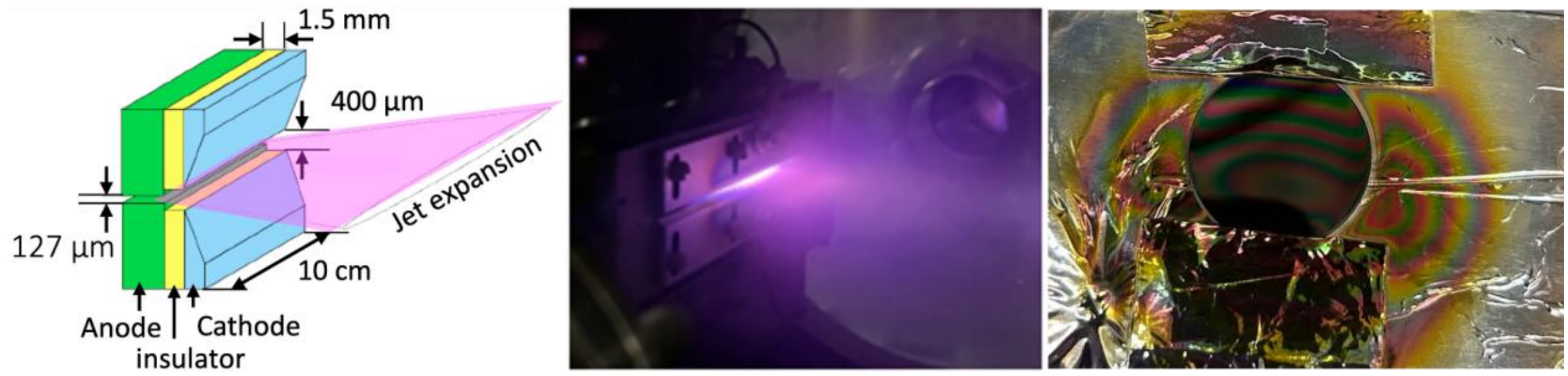

**Figure 1.** (Left) Schematic of the PDN (adapted from Broks et al. 2005). (Center) Picture of the pulsed plasma planar expansion through the slit. (Right) Picture of the *tholin* sample deposited on a silicon substrate. Iridescent features resulting from the optical interference with the thin-film deposition can be seen.

***Production of Titan Tholins***

In the study presented here, $N_2$:$CH_4$ plasma experiments were conducted using a voltage of -700V to generate the plasma discharge. *Tholins* produced in the plasma volume were jet-deposited onto a silicon substrate for 10 h in order to produce a ~800-nm thick layer of material (see Figure 1). At the end of the run, the vacuum chamber was evacuated and filled with nitrogen to minimize air exposure before collecting the *tholin* sample and transferring it into the NASA Ames Atmospheric Chemistry Laboratory (ACL), where it was placed under vacuum again before simulating benzene cloud growth.

### 2.1.2. Benzene Ice Deposition and Growth

***Experimental Setup***

To study the condensation properties of benzene ice, we utilized the cloud nucleation and growth chamber of the NASA Ames ACL (Iraci et al. 2010; Santiago-Materese et al. 2018; Dubois et al. 2021). The apparatus consists of a stainless-steel vacuum chamber (Figure 2) which is continuously evacuated with a turbomolecular pump. It can reach a base pressure of 6–7 x $10^{-8}$ Torr (9.3 x $10^{-8}$–8.0 x $10^{-8}$ mbar). For the ice condensation experiments conducted in this study, a silicon substrate was placed within a copper ring connected to a vacuum-jacketed liquid nitrogen cryostat. In the case of the *tholin* sample deposited on silicon described above, the mounting and evacuation procedures were designed to minimize exposure to room air, and it is estimated that the process took approximately 45-60 minutes. The sample was kept under vacuum overnight before conducting condensation experiments the following days to ensure removal of any water adsorbed on the sample. Although air exposure was minimized as much as possible, a thin oxidized layer of a few nanometers is likely to be present on the outermost surface of the tholin sample (Carrasco et al. 2016). However, we do not expect it to significantly impact the surface physico-chemical and bulk properties of the tholins. Kapton heaters were mounted around the end of the cold finger, and the temperature was controlled by an Omega i-Series temperature controller. Temperature measurements were taken with three K-type thermocouples (their locations are shown in Figure 2). Pressure measurements were taken with an ion gauge (Terranova controller) calibrated to a Baratron capacitance manometer (0.1 Torr full scale, 0.13 mbar). KBr windows were placed on the vacuum chamber, below and above the silicon substrate position, to enable the transmission of the infrared (IR) beam (Figure 2). Using a Thermo Fisher iS50 Fourier Transform Infrared (FTIR) spectrometer, we measured IR transmission spectra of the whole surface of the sample from 6000–650 $cm^{-1}$ (1.7–15.4 μm) in real time during deposition, to detect the appearance of benzene ice absorption bands. Spectra were taken with a spectral resolution of 1 $cm^{-1}$. The OMNIC software (version 9.15) was employed to calculate the integrated peak area of the most intense bands (see Section 2.1.3).

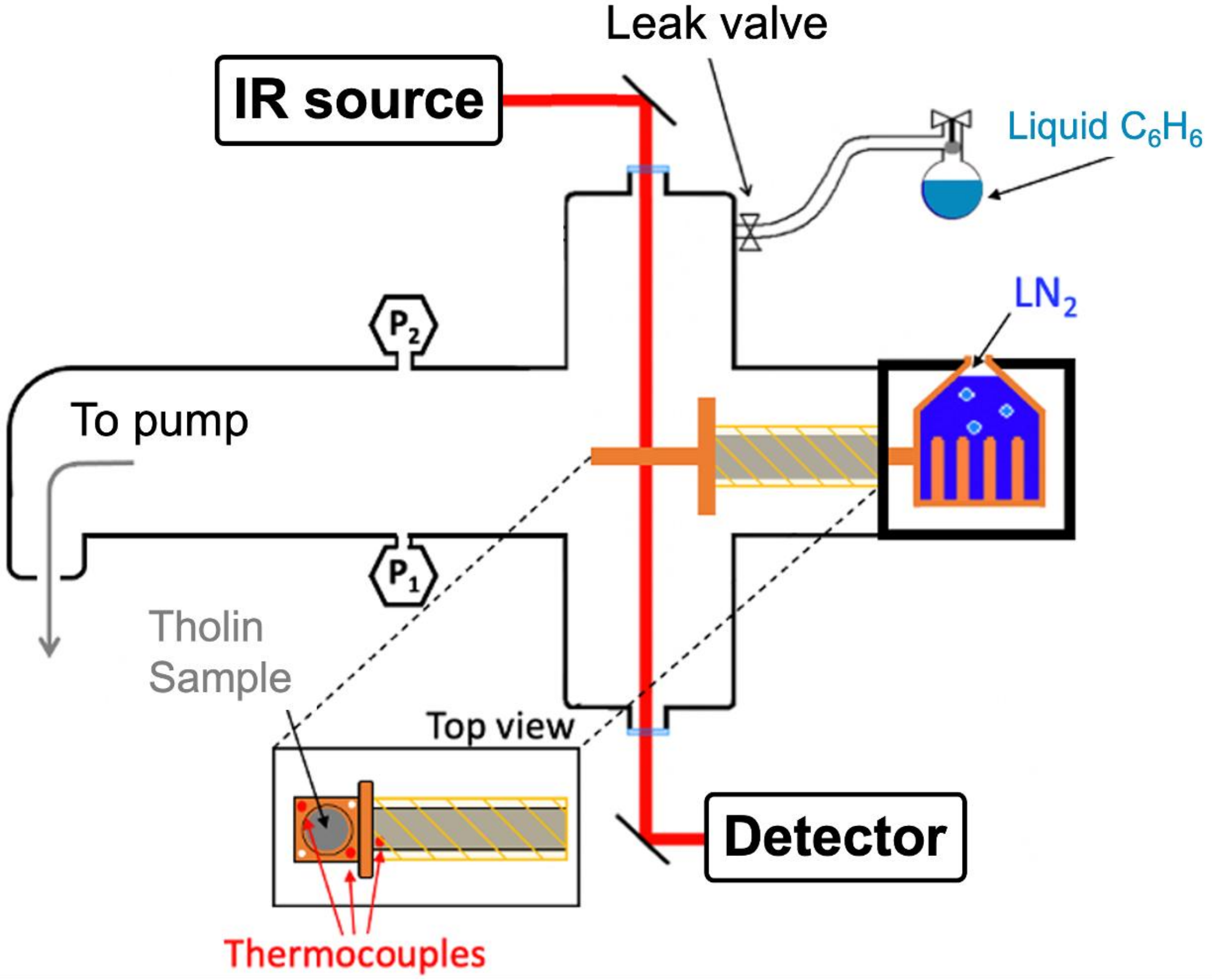

**Figure 2.** Schematic diagram (not to scale) of the experimental apparatus used for benzene ice nucleation and growth studies (adapted from Dubois et al. (2021)). Pressure was measured with a capacitance manometer (P1) and an ion gauge (P2). The inset shows a top view of the cooled sample holder with the positions of the three K-type thermocouple gauges (red dots) used for temperature measurements. The ice sample formed on either or both sides of the silicon substrate/*tholin* sample (grey), which was in the path of the IR beam of the Thermo Fisher iS50. Infrared transmission spectra were collected with a liquid $N_2$-cooled MCT/A detector.

### *Gas and Sample Preparation*

A glass bulb was prepared with 5 mL liquid benzene ($C_6H_6$, HPLC grade, > 99.9% purity). Once filled, the bulb was placed in a dry ice/ethanol slush (< -60ºC) and freeze-pumped through a glass gas handling manifold to remove any volatile impurities and residual air before being connected to the vacuum chamber via a stainless-steel delivery line. The line was then pumped down through the vacuum chamber to remove any residual air (usually for 1–2 hr). Once the base pressure was deemed satisfactory (< 1 x $10^{-7}$ Torr, or 1.3 x $10^{-7}$ mbar) and stable conditions were reached, $C_6H_6$ vapor was introduced from the bulb into the vacuum chamber, and gas flow was controlled by an ultrafine all-metal leak valve (Lesker VZLVM940R), following the protocol described in Dubois et al. (2021). This liquid sample was used for several experiments over several days. At the beginning of each day of experiments, the bulb was frozen in a dry ice and ethanol slush once more and any residual contaminants were pumped away through the vacuum chamber.

Before injecting $C_6H_6$ vapor, the *tholin* sample mounted in the chamber was first rapidly brought to a temperature below 160 K then slowly cooled (0.5 K min$^{-1}$) closer to a target temperature between 138 and 157 K. Once the target temperature was reached, $C_6H_6$ vapor was introduced into the chamber by increments of 1 x 10$^{-6}$ Torr and held at each step for 30–60 s, while monitoring the $C_6H_6$ vibrational bands between 6000–650 cm$^{-1}$ (1.6–15.4 μm) and their peak area growth rates in the acquired transmission spectrum (see Table 1). After the equilibrium phase, the temperature of the substrate was brought back up to sublimate the ice. Subsequently, this entire process was then repeated for different temperatures, enabling $S_{crit}$ measurements for nucleation of $C_6H_6$ between 138–157 K (see Table 2).

### *Pressure measurements and uncertainty analysis*

Ion gauge sensitivity factors for benzene were determined by comparison to the gas-insensitive manometer in the region where their working ranges overlap, following the protocol in Dubois et al. (2021). Three cross-calibration experiments were conducted over several days where benzene gas was injected into the chamber while increasing the pressure from ~10$^{-5}$ to 10$^{-4}$ Torr (1.3 x 10$^{-5}$ to 1.3 x 10$^{-4}$ mbar) in ~10–12 increments. Each pressure condition was held constant for 3–5 minutes while the pressure was measured with both the ion gauge and the Baratron capacitance manometer. These time-invariant conditions were achieved in a flow regime (continuous inlet and pumping of gas). These calibration experiments provided an average $C_6H_6$ pressure correction factor of 5.62 for the ion gauge, by which the recorded "raw" ion gauge pressure was divided for the analysis of subsequent equilibrium measurement data. A standard deviation among the replicate measurements for benzene of 2.6% was taken as the uncertainty in the ion gauge sensitivity factor. The overall uncertainty on pressures measured during benzene ice equilibrium experiments was determined through the square root of the sum of the squares of four contributing factors: i) the uncertainty on the gas factor used to convert the ion gauge measurement to gas-independent measurements (*i.e.*, 2.6% of the calibrated pressure); ii) the uncertainty due to the assumption that ~50% of the initial base pressure in the chamber corresponds to residual benzene vapor from a previous experiment (N.B.: this factor is generally negligible); iii) the uncertainty due to the drift in pressure during the period of time the ice is kept at equilibrium (standard deviation of the pressure over the equilibrium time period); and iv) the uncertainty due to the finite nature of the spacing between pressure steps which can be assessed using the existing apparatus and protocol (generally the largest source of uncertainty).

### *Equilibrium Temperature Calibration*

The calibrated equilibrium temperature at each of the measurement points (*T*) was calculated using the following vapor pressure equation for crystalline benzene from Hudson et al. (2022a):

$$ln(P) = (-5978.2)(1/T) + 25.6 \qquad (1)$$

where *P* is the equilibrium vapor pressure (in Torr). This vapor pressure relationship was derived from the earlier study of Dubois et al. (2021). The uncertainty on temperatures reported below are propagated from the uncertainty in the pressure measurement used to calibrate the Baratron manometer with the ion gauge.

### *Benzene ice critical saturation ($S_{crit}$) measurements*

For the $C_6H_6$, $S_{crit}$ experiments, the substrate was brought to a given target temperature that was then kept constant and the benzene partial pressure was adjusted while monitoring the IR spectrum through the silicon substrate to detect the onset and growth of $C_6H_6$ ice. Infrared peak areas were integrated for each of the four most intense bands corresponding to the $\nu_{11}$ C–H bending mode between 670–710 cm$^{-1}$, the $\nu_{18}$ twist mode between 960–1050 cm$^{-1}$, the $\nu_{19}$ C–C stretching modes between 1450–1500 cm$^{-1}$, and the $\nu_2 + \nu_{20}$ combined C–H stretching modes between 2950–3100 cm$^{-1}$. These bands, their assignments, and the corresponding references are listed in Table 1, following the Wilson notation (Wilson et al. 1955; Bernhardsson et al., 2000) for the band assignments, and the Herzberg notation when indicated in parenthesis. As the strongest absorption band for $C_6H_6$ ice in the accessible spectral range, the $\nu_{11}$ band was integrated to monitor the ice onset, growth and equilibrium. This $\nu_{11}$ band is the same spectral band that was observed with CIRS by Vinatier et al. (2018) and indicated as $\nu_4$ following the Herzberg notation.

**Table 1.** Most intense IR bands observed during crystalline benzene ice experiments, their assignments and respective references. *N.B.*: Band assignments are provided in both the Wilson notation (Wilson et al. 1955) and the Herzberg convention (Herzberg 1972, in parentheses).

| Wavelength (cm$^{-1}$) | Assignment $^{a, (b)}$ | Mode | References |
|---|---|---|---|
| 670-710 | $\nu_{11}$ ($\nu_4$) | C-H bend | *c, d* |
| 960-1050 | $\nu_{18}$ ($\nu_{14}$) | twist | *c, e, f, g, h* |
| 1450 – 1500 | $\nu_{19}$ ($\nu_{13}$) | C-C stretch | *c, e, f, g, h* |
| 2950 – 3100 | $\nu_{20}$ ($\nu_{12}$) and others | C-H stretch | *c, f, g, h* |

**Note:**

*a: Wilson et al. 1955; b: Herzberg 1972; c: Bernhardsson et al. 2000; d: Vinatier et al. 2018; e: Engdahl & Nelander 1985; f: Kearley et al. 2006; g: Dawes et al. 2018; h: Chandrasekaran et al. 2011*

## 2.2. Microphysics Modeling: The Community Aerosol and Radiation Model for Atmospheres (CARMA)

Once the benzene $S_{crit}$ was measured experimentally, we implemented these new and first experimental values in CARMA simulations to assess the microphysical effects of the critical saturation and relate these to observable properties of Titan's benzene cloud. Specifically, the PlanetCARMA version, which has previously been adapted to model Titan's atmosphere (Barth 2017; Barth 2020), was used to model physical processes including coagulation, vertical transport, and cloud particle nucleation, condensation, and evaporation. These new CARMA simulations follow the ones presented in Dubois et al. (2021) but now make use of the experimentally measured $C_6H_6$ vapor pressure and critical saturation ratios (see Section 3). Further parameters relevant to the cases presented here are listed in Section 4.

# 3. Results

Figure 3 shows a mid-IR absorption spectrum, taken between 6000–650 $cm^{-1}$ (1.7–15.4 µm), of $C_6H_6$ ice deposited at T = 156.8 K onto Titan *tholins* produced in COSmIC by plasma chemistry in a $N_2$:$CH_4$ (95:5) gas mixture. The spectrum displays typical absorption features from the *tholin* sample. In the 1200–1800 $cm^{-1}$ region a prominent broad feature characteristic of n–$CH_3$ and n–$CH_2$ bending modes, as well as C=N, C=C, and (hetero)aromatic signatures are observed. The double-feature observed between 2000–2300 $cm^{-1}$ corresponds to nitrile –CN and isocyanide –NC stretching modes. The relative absorbance is almost as high as that of the aliphatic and aromatic aforementioned bands, indicative of a nitrogen-rich sample. –$CH_2$ and –$CH_3$ asymmetric stretching modes often found in Titan *tholins* around 2930 and 2960 $cm^{-1}$ are not notably detected. However, the broad band seen at 3100–3600 $cm^{-1}$ results from the presence of primary and secondary amine, –NH and –$NH_2$, respectively. The relative intensities of these absorption bands all highlight a strong nitrogen incorporation into the *tholins*.

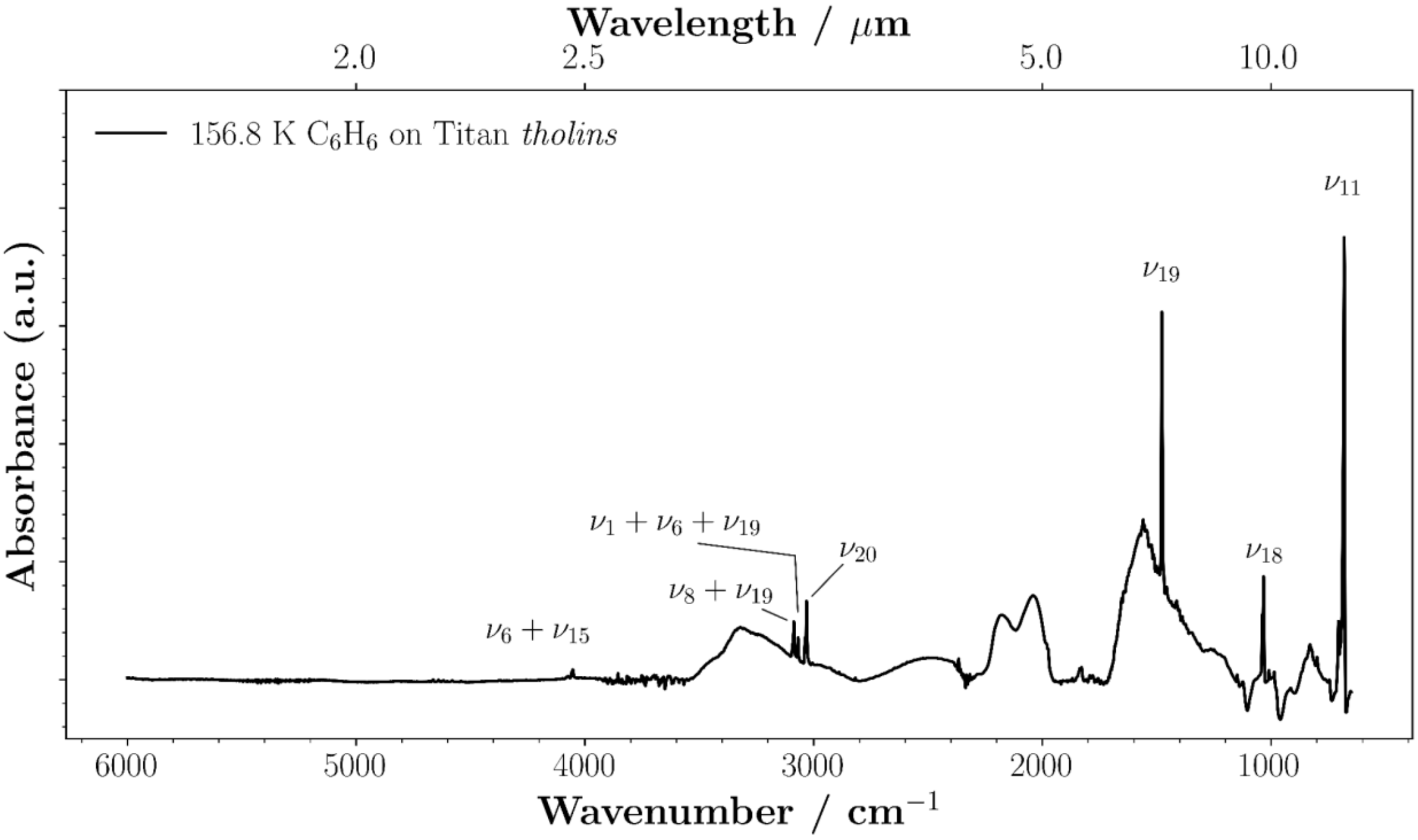

**Figure 3.** IR absorption spectrum of crystalline $C_6H_6$ ice deposited on Titan *tholin* sample at 156.8 K. Major absorption bands for $C_6H_6$ ice are indicated with the Wilson notation. The $\nu_6 + \nu_{15}$ combination mode follows the Herzberg notation as it is the only notation available. Atmospheric correction was applied to the spectrum to remove features due to residual carbon dioxide and water vapor in the portion of the beam path outside the vacuum chamber.

Sharp crystalline benzene features are seen at 682 $cm^{-1}$ (C–H bending mode, $\nu_{11}$), 1025 $cm^{-1}$ (twist, $\nu_{18}$), 1475 $cm^{-1}$ (C–C stretching mode, $\nu_{19}$), and the smaller 3025–3050 $cm^{-1}$ (combined C–H stretching mode, $\nu_{20} + \nu_2$) bands. Smaller crystalline features can also be seen at 4050 $cm^{-1}$, previously reported in crystalline 100 K benzene ices measured by Hudson et al. (2022b).

For each experiment, the integrated peak areas, sample temperature, and the vapor pressure were measured continuously to monitor ice nucleation onset ($P_{obs/C6H6}$), vapor/ice equilibrium ($P_{equ/C6H6}$), and ice sublimation. An example of these measurements is shown in Figure 4. This particular measurement shows that ice onset began after 20 min (P = 5 x $10^{-6}$ Torr) and ice was allowed to grow for 10 min before conditions were adjusted to reach vapor/ice equilibrium. After maintaining these equilibrium conditions for ~10–15 min, ice sublimation was initiated (after 63 min) by increasing the sample temperature.

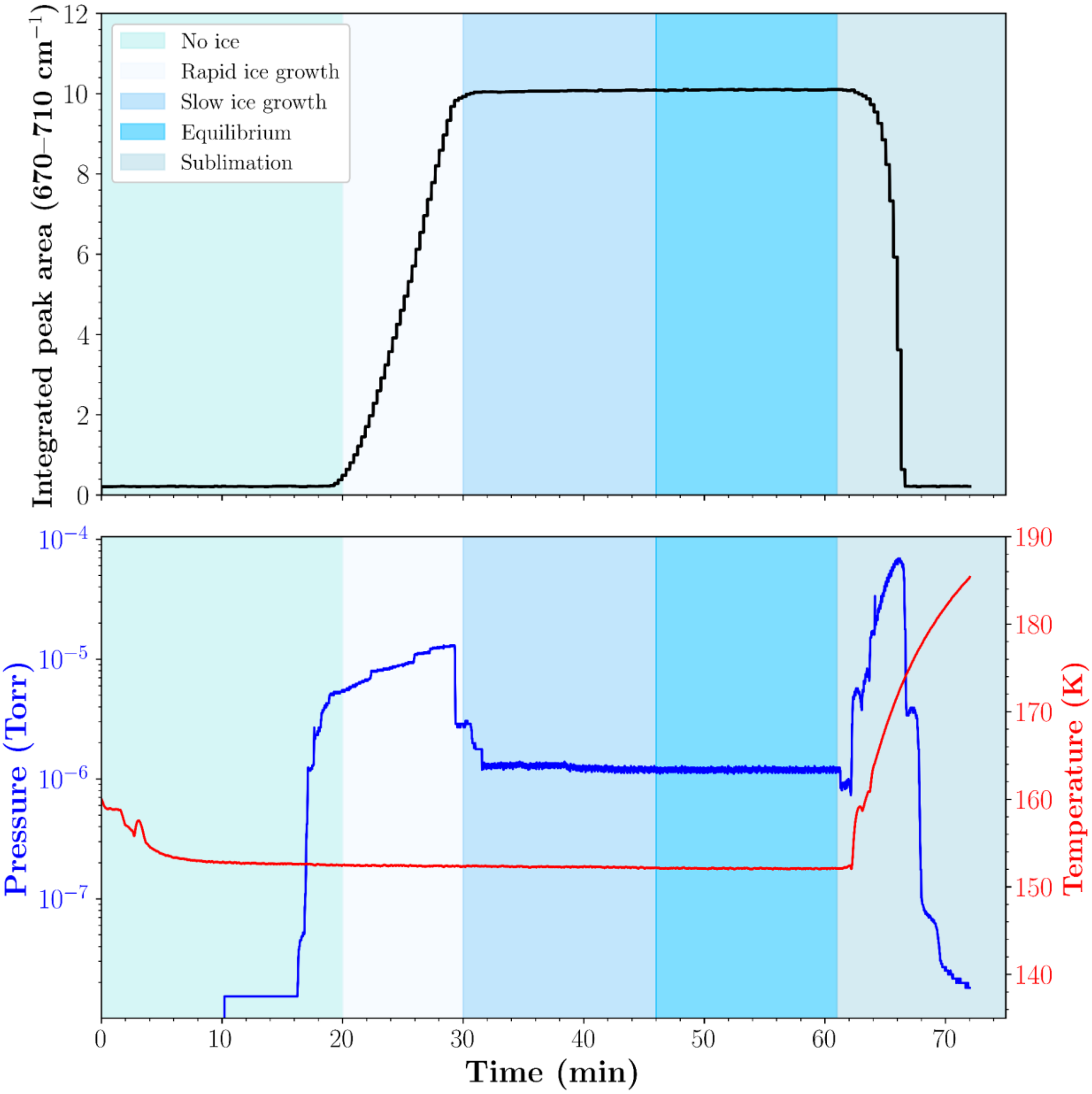

**Figure 4.** (Top) Integrated band area of the C–H bending mode between 670–710 $cm^{-1}$ as a function of time during a typical $S_{crit}$ experiment, here shown at T = 156.8 K. The shaded areas represent the succession of steps needed to grow ice and reach equilibrium conditions. (Bottom) Pressure (Torr) and sample temperature (K). Benzene vapor was first introduced at 16 min and increased in steps approximately every 1–4 min. At 30 min, the flow rate was reduced to enable the determination of equilibrium conditions.

The saturation ratio for nucleation $S_{crit}$ for each experimental temperature could then be calculated from:

$$S_{crit} = \frac{Pobs/C6H6}{Pequ/C6H6} \qquad (2)$$

where $P_{obs/C6H6}$ is the observed measured vapor pressure at nucleation and $P_{equ/C6H6}$ is the equilibrium vapor pressure using Eq. (1) at the same temperature (Iraci et al. 2010). Following the protocol described in section 2.1.2 and Dubois et al. (2021), an offset was applied to the measured temperature and uncertainties were calculated from the square-root of the sum of the squares of the four main uncertainty contributions for each experimental point (see Section 2.1.2 and Dubois et al. 2021 for protocol details). The resulting $S_{crit}$ values as a function of temperature are shown in Figure 5 and listed in Table 2.

On silicon, the largest $S_{crit}$ value of 29.8 ± 7.3 was obtained for a nucleation temperature of 139.3 ± 0.8 K (Table 1). On the *tholin* sample, the $S_{crit}$ value measured at a nucleation temperature of 138.2 ± 0.8 K was noticeably lower, at 13.8 ± 6.6. We interpret the coldest point at 139.1 ± 0.8 K on silicon ($S_{crit}$ = 13.6) to be an experimental outlier, possibly due to inaccuracies in pressure stabilization and base pressure readings, as this was the last measurement performed at the end of a long series conducted all on one day. The smallest $S_{crit}$ value on silicon of 11.0 ± 1.6 is measured at a nucleation temperature of 153.3 ± 0.6 K, whereas it is 3.3 ± 0.6 for nucleation on *tholin* at a temperature of 156.8 ± 0.7 K.

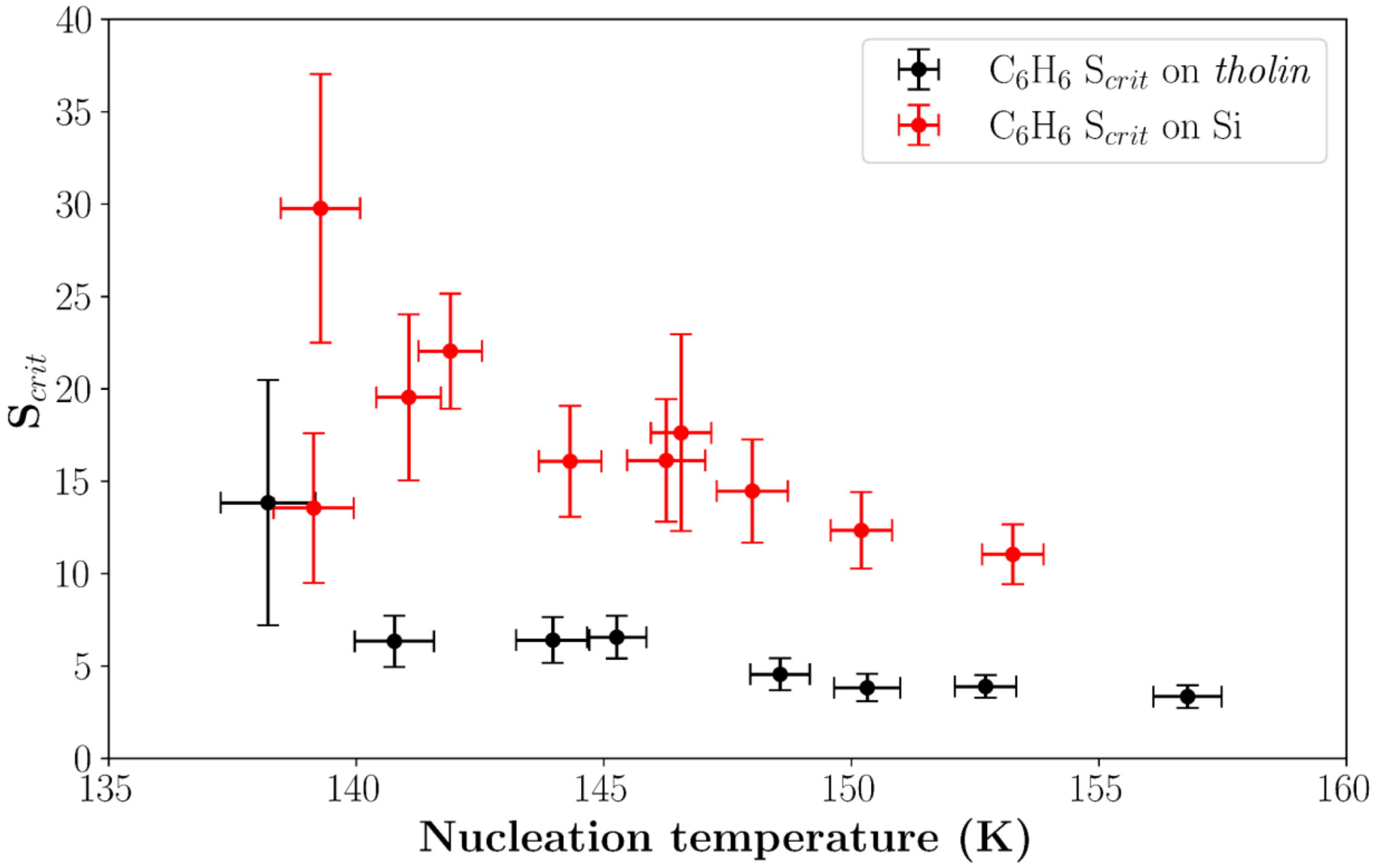

**Figure 5.** $S_{crit}$ values measured for $C_6H_6$ between 138–157 K on a bare silicon substrate (red) and on a laboratory-produced Titan *tholin* sample (black). Uncertainties in temperature and $S_{crit}$ calculations are also shown (see Methods).

These measurements for benzene nucleation conducted on bare silicon and Titan *tholin* show a clear influence of the Titan *tholin* on the $C_6H_6$ nucleation, where $S_{crit}$ values for benzene ice on the *tholin* sample (black) are much lower than on the blank silicon substrate used as a reference (red), by a factor of ~3–5 over the studied temperature range. This indicates that a lower degree of supersaturation is necessary for nucleation of $C_6H_6$ to proceed on the Titan *tholin* sample than on the bare silicon substrate. Thus, the Titan *tholin* substantially favors the condensation and nucleation of $C_6H_6$. Additionally, our nucleation measurements on both the *tholin* sample and Si substrates over the 138–157 K temperature range allow us to derive a temperature dependence to $S_{crit}$, which increases with decreasing temperature. This is supported by the general observation that colder temperatures require higher saturation ratios (as seen in Iraci et al. 2010 and references therein). This inverse relation where $S_{crit}$ values are higher at lower temperatures is not predicted using classical nucleation theory, whereby the contact parameter is assumed to be constant. Instead, the thermodynamic driving force (*i.e.*, the chemical potential difference between the vapor and the ice) competes with the molecule's kinetic barrier. With decreasing temperature, the molecular thermal velocity decreases and the residence time of the adsorbed molecules on the rough surface (*i.e.*, *tholins*) decreases exponentially relative to the diffusion time (see Eq. 3). As such, the net condensation flux decreases because the desorption flux does not drop as fast. Therefore, to maintain the net growth rate needed to overcome the critical free energy $\Delta F_g$, the vapor pressure must be raised significantly, and the homogeneous nucleation limit is approached. We note that this effect was also seen with the condensation of water on sodium chloride and sodium perchlorate substrates (Santiago-Materese et al. 2018).

**Table 2.** Results for $S_{crit}$ $C_6H_6$ ice laboratory measurements on bare silicon and *tholin* substrates with their associated experimental uncertainties.

| **ID** | **T nucleation (K)** | | **$S_{crit}$** | |
|---|---|---|---|---|
| | Silicon | Tholin | Silicon | Tholin |
| a | 139.1 ± 0.8 | 138.2 ± 1.0 | 13.5 ± 4.1 | 13.8 ± 6.6 |
| b | 139.3 ± 0.8 | - | 29.8 ± 7.3 | - |
| c | 141.1 ± 0.7 | 140.8 ± 0.8 | 19.5 ± 4.5 | 6.3 ± 1.4 |
| d | 141.9 ± 0.6 | - | 22.0 ± 3.1 | - |
| e | 144.3 ± 0.6 | 144.0 ± 0.7 | 16.1 ± 3.0 | 6.4 ± 1.2 |
| f | 146.3 ± 0.8 | 145.3 ± 0.6 | 16.1 ± 3.3 | 6.6 ± 1.2 |
| g | 146.6 ± 0.6 | - | 17.6 ± 5.3 | - |
| h | 148.0 ± 0.7 | 148.6 ± 0.6 | 14.5 ± 2.8 | 4.5 ± 0.9 |
| i | 150.2 ± 0.6 | 150.3 ± 0.7 | 12.3 ± 2.1 | 3.8 ± 0.7 |
| j | 153.3 ± 0.6 | 152.7 ± 0.6 | 11.0 ± 1.6 | 3.9 ± 0.6 |
| k | - | 156.8 ± 0.7 | - | 3.3 ± 0.6 |

# 4. Discussion

## 4.1 Impact of *tholins* on benzene ice condensation

By measuring the $S_{crit}$ of benzene on tholins compared to bare silicon, we observe a significant impact of *tholins* on benzene ice nucleation. The much lower $S_{crit}$ values shown for nucleation on *tholins* (black data in Figure 5) indicate that heterogeneous nucleation of benzene clouds on *tholins* will be significantly easier than on any surfaces similar to that of the blank silicon substrate.

Furthermore, our results show that a single $S_{crit}$ value cannot be representative of a wide temperature range, consistent with previous findings for water ice nucleation on a variety of surfaces (*e.g.*, Iraci et al. 2010). This had not been previously confirmed for benzene as the condensing species. This temperature dependence (a factor of 4 from 3.34 to 13.83 as nucleation temperature is reduced from 148.0 K to 139.1 K) should asymptotically reach a value of unity for temperatures >>148 K, as discussed in Iraci et al. (2010).

In light of these two factors, we recommend the use of the values presented here, in this temperature regime, particularly since Titan *tholins* represent good analogues for Titan's haze particles (Sciamma-O'Brien et al. 2023) and act as efficient condensation nuclei. The *tholins* produced for this study form micrometric-sized fractal aggregates of smaller-sized nanoparticles deposited as a thin layer which is 100s of nm thick (Sciamma-O'Brien et al. 2017). These particles present favorable temperature-dependent nucleation sites where nucleation and growth might differ from predictions of classical nucleation theory (Rapp and Thomas, 2006). Our results follow a similar trend to previous $S_{crit}$ studies of water ice (with an inverse relation between $S_{crit}$ and temperature), and further confirm the temperature- and substrate-dependency of the nucleation properties of atmospheric ices (Knopf & Koop, 2006; Trainer et al. 2009; Iraci et al. 2010; Phebus et al. 2011).

## 4.2 Implications of the temperature dependence on contact parameters used in microphysical simulations

In order to relate our new laboratory measurements to observable properties of Titan's south polar cloud, we ran microphysics simulations using CARMA. The model column covered 30–400 km in 1-km thick layers. Simulations were run using the $C_6H_6$ abundance and temperature profile for 87° S derived by Vinatier et al. (2018), *i.e.*, fixing a volume mixing ratio (vmr) for $C_6H_6$ of $1.8 \times 10^{-6}$ at 380 km (see below). Moreover, our simulation also included a downwelling circulation at the south pole observed/expected at this season on Titan. More details regarding how CARMA treats the descending branch of Titan's circulation and its physical parameters can be found in section 3.3 of Dubois et al. (2021).

In CARMA, the measured critical saturation values are converted to a contact parameter in the nucleation rate equation. Classical nucleation theory is used to calculate the nucleation rate of benzene ice particles,

$$J = K \exp( -\Delta F_g / k_B T ), \qquad (3)$$

where $\Delta F_g$ is the free energy of germ formation, $k_b$ is Boltzmann's constant, $T$ is the temperature and the prefactor $K = \pi Z_s a_g^2 \beta c_1$. Terms in the prefactor $K$, *i.e.*, the Zeldovich factor ($Z_s$), flux of monomers ($\beta$), and surface monomer concentration ($c_1$), are detailed in Barth & Toon (2003). The energy term,

$$\Delta F_g = 4/3\ \pi\ a_g^2\ \sigma_{sv}\ f(m), \qquad (4)$$

dominates the nucleation rate and is a function of the critical germ radius, $a_g$, and the surface energy, $\sigma_{sv}$, in our case, for the solid-vapor transition of benzene. The effect of the contact parameter, $m$, is included in the function $f(m)$. The contact parameter represents the interaction between the condensing species and condensation nuclei in heterogeneous nucleation. Values range from -1 to 1, where $m = 1$ indicates there is no energy barrier to nucleation and $m = -1$ is equivalent to homogenous nucleation. We followed Barth & Toon (2003) description of the nucleation rate equation and their method to derive the contact parameter from the critical saturation and temperature measurements. To summarize, we iteratively solved for $m$ in the nucleation rate equation for each of the measured $S_{crit}$ and $T$ combinations shown in Figure 5. The resulting points are shown in Figure 6. We then used orthogonal distance regression to fit a line to the points (including error). For the silicon data set, this line is nearly flat indicating little variation in the contact parameter with temperature. For the *tholin* data set, we find $m$ =

2.6e-4 $T$ + 0.9494. We note that when deriving the contact parameter, a number of other parameters, provided in Table 3, are included in the nucleation rate equation. After the contact parameter and the vapor pressure, the surface energy has the largest effect on the nucleation rate.

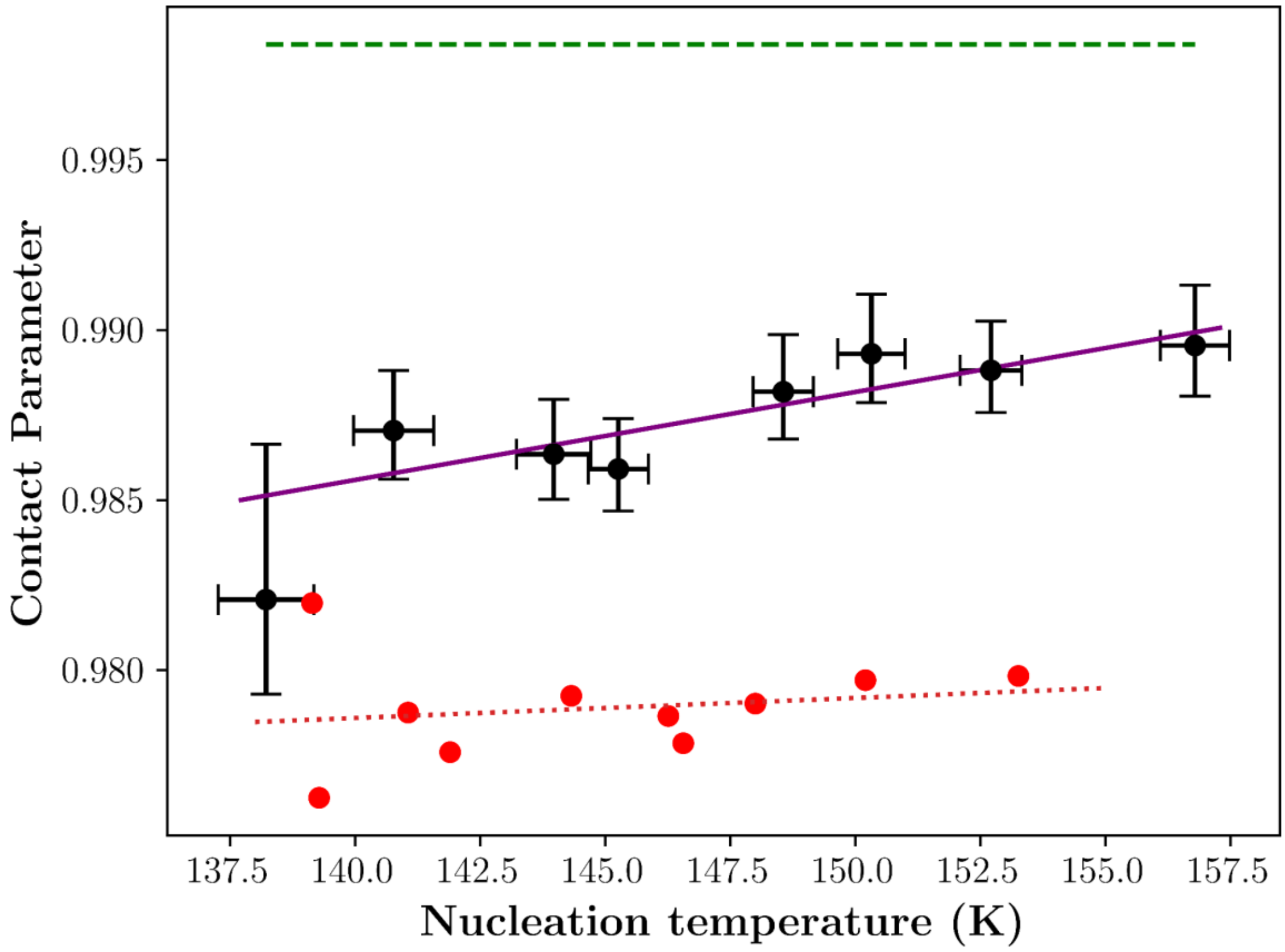

**Figure 6.** Contact parameter values calculated from the laboratory measurements of critical saturation as a function of temperature for $C_6H_6$ nucleation onto *tholins* (black points and error bars) and linear fit (purple line). Shown in red are the contact parameters and linear fit to the measurements of benzene nucleation on a silicon substrate. The green dashed line shows the temperature-independent contact parameter used for the benzene study in Dubois et al. (2021).

**Table 3.** $C_6H_6$ nucleation rate parameters considered for Titan's atmosphere. Additional comments related to the assumptions for the main physical constraints are also indicated.

| Parameter | Ref/Eqn/Value | Comments | Used in |
|---|---|---|---|
| Contact parameter | $m$ = 2.6e-4 $T$ + 0.9494 (see Fig. 6) | Determined from the $S_{crit}$ values in this paper (Fig. 5) | $\Delta F_g$ |
| Equilibrium vapor pressure | Hudson et al. (2022a) | Applicable to cold Titan temperatures | $a_g$, β ($\Delta F_g$, K) |
| Surface energy ice-air | $\sigma_{iv} = \sigma_{il} + \sigma_{lv}$ | Antonoff's rule | $\Delta F_g$, $a_g$ |
| Surface energy ice-liquid | Curtis et al. (2005) | $\sigma_f = 0.32$ | $\sigma_{iv}$ |
| Surface energy liquid-air | Jasper (1972), Table 23 | liquid benzene and $N_2$+$O_2$ vapor | $\sigma_{iv}$ |
| CCN particle density | 1000 kg/m$^3$ | Standard value for *tholins* | $\Delta F_g$, $a_g$ |
| $C_6H_6$ ice density | 1015.3 kg/m$^3$ | Richards et al. (1921); T=273 K | $\sigma_{il}$ |
| $C_6H_6$ monomer concentration | $c_1 = 10^{19}$ m$^{-2}$ | Assume monolayer on surface | $K$ |

The condensation curve for benzene at Titan's south pole indicates saturation (S = 1) occurs at an altitude of 280 km. In Dubois et al. (2021) we assumed a smaller critical saturation ($S_{crit}$ = 1.35) and found cloud particles formed closer to an altitude of 280 km. However, from our experimental $S_{crit}$ measurements, significant nucleation is expected to be inhibited until much larger saturation values are reached (*e.g.*, Table 2 where $S_{crit} > 3$). To show the important influence of the nucleation critical saturation on the height of the cloud tops, we ran microphysics simulations with CARMA using different nucleation contact parameters (Figure 7): a temperature-dependent contact parameter calculated from a linear fit to the Titan tholin $S_{crit}$ (black curve) ; a constant contact parameter $m$ = 0.9821 calculated from a unique $S_{crit}$ value on tholins at the lowest measured temperature T = 138.22 K (blue curve); a contact parameter calculated from a fit to all $S_{crit}$ measured on bare silicon (red dotted curve); and an approximated contact parameter based on a butane analog that was used in the Dubois et al. (2021) study (green dashed curve). We found that cloud particles form more readily in the nominal simulation (black curve) at the ~260 km altitude level, where T = 125 K and m = 0.9819. The $87^{o}$ S temperature profile is also shown in Figure 7. The benzene mixing ratio is bounded by a flux at the top boundary and gas is permitted to diffuse across the bottom boundary. The altitude region where the vmr curves diverge reflects the relative differences in the nucleation rate between the four model runs. The loss of more vapor in the curve obtained with the $S_{crit}$ value used in Dubois et al. (2021) is also indicative of the larger number of cloud particles to serve as sites for benzene condensation. Below ~230 km, condensation is the dominant process to remove benzene from the atmosphere; condensation occurs for $S\cong1$ (with a slight increase due to the Kelvin effect) and so the critical saturation has no effect. Because $S_{crit}$ is significantly above 1.0 for $C_6H_6$ in Titan's atmosphere, the environment has a large degree of supersaturation when cloud particles do form. There is more benzene available for growth through condensation. The peak in the relative humidity is at z ~ 250 km. The largest nucleation peak is seen in the size distribution curve for this altitude and a smaller growth mode peak is seen as well. The rate of condensation growth is proportional to the vapor pressure and, as the vapor pressure increases with increasing temperature, the condensation growth rate increases. The warmer temperatures at 260 and 270 km explain seeing only the growth mode peak in the size distribution curves here. The relative humidity drops significantly below 250 km, down to nearly 100% at 230 km to the model bottom boundary at 30 km. The size distribution remains invariant below 230 km.

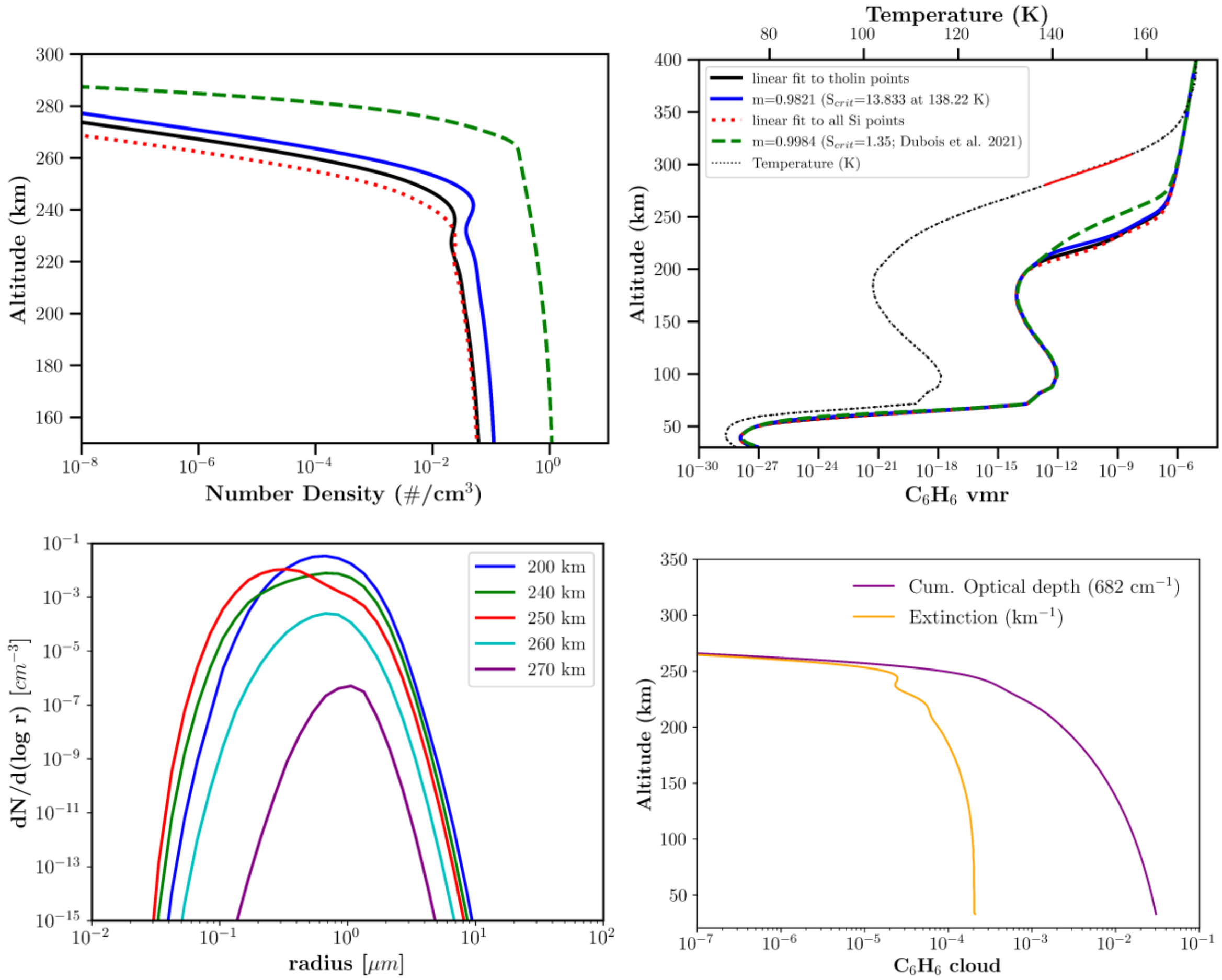

**Figure 7.** Cloud particle number density (top left) and gas volume mixing ratio (vmr) profiles (top right) for model runs with different nucleation contact parameters. The 87° S temperature profile is shown as a black dotted line, and the portion relevant to our experimental temperature range (138 to 157 K) is indicated in red. Solid lines correspond to models run with the contact parameters calculated from the *tholin* laboratory measurements (linear fit (black) and lowest temperature measurement (blue)). For comparison, the red dotted line shows the model results when using the contact parameter calculated from the fit to all silicon $S_{crit}$ measurements, and the green dashed line corresponds to model results obtained with contact parameter approximations based on a butane analog before the current benzene nucleation measurements were available (Dubois et al. 2021). (Bottom left) Calculated particle size distribution for 5 altitude levels obtained with the "linear fit" contact parameter. (Bottom right) Calculated cumulative optical depth and extinction (lower right) obtained from the black number density curve (*i.e.*, with the linear fit contact parameter), and the $C_6H_6$ refractive index of Schmitt et al. (2015) measured between 400 – 4199 $cm^{-1}$.

Microphysics models, particularly when extended beyond terrestrial applications, rely on parameters that are often poorly constrained (*e.g.*, Table 3). By making laboratory measurements of these parameters for the relevant temperature-pressure regime, as we have done for benzene cloud formation in Titan's atmosphere, we are able to create more physical simulations to better address the observations. The previous use of a single contact parameter value without a temperature dependence is shown to be an oversimplification. With these relevant laboratory observations, there is no longer a need to rely on the extrapolated butane analog used previously, which produced values much larger than the contact parameters derived here for benzene ice condensed on *tholins*. Our modeling has shown that the saturation vapor pressure and critical saturation have the most significant effect on the cloud top height. Using the newly measured values of these parameters, our modeled cloud formation produces an observable cloud top approximately 20 km lower than indicated from previous analysis of Cassini data (Vinatier et al. 2018). The lower cloud tops we have modeled here may be an indication of the presence of co-condensation which would alter the saturation vapor pressure and critical saturation of the mixture, and allow future models to reproduce the cloud formation closer to 280 km at the south pole.

## 5. Conclusions

Titan's circulation dynamics underwent a global change following the northern hemisphere spring equinox in 2009. The descending branch of the global circulation cell observed in the south pole in 2010 resulted in stratopause heating (at ~400 km) and an enrichment of haze at high southern latitude. The CIRS instrument onboard Cassini revealed the presence of several organic ices in the large stratospheric cloud of Titan's south pole following this circulation reversal. Among these organics, benzene was for the first time clearly identified by Vinatier et al. (2018). However, a full understanding of the altitude, microphysics, and the possibility for condensation of benzene ice has remained elusive. The microphysical interpretation of these observations depends largely on the low-temperature properties of these ices, given the cold temperatures found at the formation altitude of these clouds.

In order to study the low-temperature nucleation of $C_6H_6$, we performed experimental measurements of the critical saturation ratios ($S_{crit}$) for $C_6H_6$ ices between 138–157 K, on bare silicon and Titan *tholin*. Our measurements on bare silicon provide a benchmark for benzene $S_{crit}$, while ice deposition onto *tholins* represents more realistic conditions of benzene condensation and growth on the solid particles composing Titan's haze. As shown in Figure 5, there is a considerable temperature dependence to the onset conditions required for benzene ice nucleation on *tholin* materials, with $S_{crit}$ values greater than 5 when T < ~145 K.

Based on these new experimental results, we also conducted a new analysis of the south Polar cloud microphysics with CARMA. From our new $S_{crit}$ measurements, we calculated a temperature-dependent nucleation contact parameter of $m = 2.6e\text{-}4\ T + 0.9494$ that was used in CARMA to derive the benzene mixing ratios and the cloud number density at 87º S. Our work has shown that this experimentally-derived contact parameter had the most significant impact on the cloud top height, compared to the influence of other model variables. Our measured $S_{crit}$ values result in modeled cloud formation producing a benzene-on-aerosols observable cloud top approximately 20 km lower than determined from a previous analysis of Cassini data (Vinatier et al. 2018). Thus, our results suggest that the cloud studied by Vinatier et al. (2018) was likely influenced by co-condensation of more than one volatile gas species. Based on previous modeling (Barth 2017), propionitrile $C_2H_5CN$ may condense 10-15 km above $C_6H_6$. Similarly to other species condensing in this colder polar region, it is also possible that the condensation altitude of nitriles such as propionitrile would also drop, thus facilitating the co-condensation of $C_6H_6$ and $C_2H_5CN$. However, acrylonitrile $C_2H_3CN$ was shown by Vinatier et al. (2018) to be a likely candidate for the other unidentified signatures centered at 695 $cm^{-1}$, although these were detected at lower altitudes. Nevertheless, co-condensation would alter the saturation vapor pressure and critical saturation of the mixture, and allow for cloud formation at higher altitudes in the south pole.

## Acknowledgements

The authors thank two anonymous reviewers whose comments helped to meaningfully improve the quality of the manuscript. This research was supported by the Cassini Data Analysis Program of NASA’s Science Mission Directorate, Project 80NSSC23K0968. The authors acknowledge and appreciate technical support from Emmett Quigley and Richard Kolyer. Credit is due to Lauren Scovel for the plasma photographs.